\documentclass[prd,floatfix]{revtex4}

\usepackage{amsmath,amsfonts}
\usepackage[dvips]{graphicx}

\begin{document}

\title{Cosmological model with macroscopic spin fluid}

\author{Marek Szyd{\l}owski}

\affiliation{Astronomical Observatory, Jagiellonian University, 
Orla 171, 30-244 Krak{\'o}w, Poland}

\author{Adam Krawiec}

\affiliation{Institute of Public Affairs, Jagiellonian University,
Rynek G{\l}{\'o}wny 8, 31-042 Krak{\'o}w, Poland}

\begin{abstract}
We consider a Friedmann-Robertson-Walker cosmological model with some exotic 
perfect fluid with spin known as the Weyssenhoff fluid. The possibility that 
the dark energy may be described in part by the Weyssenhoff fluid is discussed. 
The observational constraint coming from supernovae type Ia observations is 
established. This result indicates that, whereas the cosmological constant is 
still needed to explain current observations, the model with spin fluid is 
admissible. For high redshifts $z > 1$ the differences between the model with 
spin fluid and the cold dark matter model with a cosmological constant become 
detectable observationally for the flat case with $\Omega_{\text{m},0}=0.3$. 
From the maximum likelihood method we obtain the value of $\Omega_{\text{s},0} 
= 0.004 \pm 0.016$. This gives us the limit $\Omega_{\text{s},0} > -0.012$ at 
the $1\sigma$ level. While the model with ``brane effects'' is preferred by 
the supernovae Ia data, the model with spin fluid is statistically admissible. 
For comparison, the limit on the spin fluid coming from cosmic microwave 
background anisotropies is also obtained. The uncertainties in the location of 
a first peak give the interval $-1.4 \times 10^{-10} < \Omega_{\text{s},0} 
< -10^{-10}$. From big bang nucleosynthesis we obtain the strongest limit 
$\Omega_{\text{s},0} \gtrsim -10^{-20}$. The interconnection between the model 
considered and brane models is also pointed out. 
\end{abstract}

\maketitle

\section{Introduction}

In 1923 Cartan introduced the intrinsic angular momentum in the theory of 
relativity (as a classical quantity) \cite{Cartan23}, before it was introduced 
as spin into quantum theory by Goudsmit and Uhlenbeck in 1925. 
The classical spin can be introduced in general relativity in two
distinct ways. The first one is to introduce spin as a dynamical quantity 
in special and then in general relativity without changing the geometry, 
i.e., without modifying the metric of spacetime \cite{Mathisson37,Honl37,%
Honl39,Honl40,Weyssenhoff47}. The spin introduced in this way showed more 
or less a similarity to the spin of quantum mechanics (and the Dirac theory 
of the electron). The second, more satisfactory way to introduce the intrinsic 
angular momentum is by generalizing the structure of spacetime. 
It was done by Cartan by assuming the metric and the non-symmetric affine 
connection as independent quantities and was further developed by Hehl 
\cite{Hehl74}, Trautman \cite{Trautman72} and Kopczynski 
\cite{Kopczynski72,Kopczynski73}. This assumption allowed a definition of the 
torsion of spacetime and its connection with the torsion with spin. In the 
framework of Einstein-Cartan theory for the Friedmann universe Trautman 
\cite{Trautman73} made the conclusion that torsion avoids the singularity 
and stops the collapse in closed models at the moment of minimum radius about 
1 cm with matter density $\rho \sim 10^{55} \mathrm{g} \, \mathrm{cm}^{-3}$ 
\[
R_{\mathrm{min}} = \left( \frac{3 G \hbar^2 n}{8 m c^4} \right)^{1/3}, 
\quad \rho_{\mathrm{max}} = \frac{4 m^2 c^2}{3 \pi^2 G \hbar^2}. 
\]

Let us note that the above formulas are valid for chaotic spin distribution 
\cite{Ponomarev}. The effects of spin fluid are important in a low-energy 
limit of the superstring theory (see for general discussion \cite{Tassie86}) 
which is supergravity whose integral part is torsion. 

The torsion contributes to the energy-momentum of spin fluid which has 
the form \cite{Ivanenko85} 
\[
T_{\mu \nu}^{\mathrm{eff}} = 
u_{\mu} u_{\nu}(p + \rho - 2s^2) 
- g_{\mu \nu}(p - s^2)
\]
with $s^2 = s_{\mu \nu} s^{\mu \nu}$, where the spin $s$ leads to 
an effective negative pressure and eliminates the singularity. 

Let us consider a world model with the Robertson-Walker symmetries which is  
filled with `perfect fluid with spin'. As it is well known \cite{Halbwachs60} 
the macroscopic spin tensor $\tau_{\mu \nu}^{\lambda}$ may be expressed in 
terms of the spin density tensor $S_{\mu \nu}$ and four-velocity of the fluid
$u^{\mu}$ ($u_{\mu} u^{\mu} = -1$)
\begin{equation}
\tau_{\mu \nu}^{\lambda} = u^{\lambda} S_{\mu \nu}, \qquad 
S_{\mu \nu} u^{\mu} = 0. 
\label{eq:1}
\end{equation}

To describe the material content of the considered model we use the 
hydrodynamical description in terms of the energy-momentum tensor which in 
the general relativity limit reduces to perfect fluid characterized by 
the energy density $\epsilon$ and the isotropic pressure $p$. In analogy, 
the physical content of the model in the Einstein-Cartan theory, based on 
the classical description of spin with equation (\ref{eq:1}), may be called 
``perfect fluid with spin''. This is surely the simplest type of 
hydrodynamic continuum of use for our aims. It is an extension of the 
well known semi-classical model of spin fluid from special relativity; 
we call it the Weyssenhoff fluid \cite{Weyssenhoff47}.
The influence of the macroscopic spin present in the fluid on the dynamics 
of the universe is then described by contributions to the energy density 
and pressure
\begin{equation}
\epsilon_{\text{eff}} = \epsilon - \frac{1}{4} S^2 , \qquad 
p_{\text{eff}} = p - \frac{1}{4} S^2 ,
\label{eq:2}
\end{equation}
where $S^2 = \frac{1}{2} S_{\mu \nu} S^{\mu \nu}$ , and $S_{\mu \nu}$  is 
the spin density tensor \cite{Arkuszewski74}. 

Supernovae type Ia observations \cite{Garnavich98,Perlmutter97,Riess98,%
Schmidt98,Perlmutter98,Perlmutter99} indicated that the Universe's expansion 
has started to accelerate during recent cosmological times. These 
observations (as well as cosmic microwave background observations) suggest 
that the energy density of the Universe is dominated by a dark energy 
component with negative pressure, driving the acceleration. The most natural 
candidate to represent dark energy is the cosmological constant. However, 
it is necessary to require a fine tuning of 120 orders of magnitude in order 
to obtain agreement with observations \cite{Peebles03}. 

In this work, we discuss the possibility that some part of the dark energy may 
be represented by the perfect fluid with spin, which is characterized by an 
equation of state in form (\ref{eq:2}). 

Let us consider the dust fluid ($p = 0$) of particles with spin $1/2$ and 
mass $m$. The energy density $\epsilon$ and absolute value of the spin 
density $S$ depend on number $n$ of particles in a unit volume
\[
\epsilon = nm, \qquad S = \frac{\hbar}{2} n.
\]
Hence we have
\begin{equation}
\epsilon_{\text{eff}} = \epsilon - \left( \frac{\hbar}{4m} \right)^{2} 
\epsilon^{2}, \qquad p_{\text{eff}} = 0 - \left( \frac{\hbar}{4m} \right)^{2} 
\epsilon^{2}.
\label{eq:3}
\end{equation}
Therefore, effective pressure is negative. Because the number of particles 
in comoving volume $n(t) \propto a^{-3}(t)$, where $a$ is the scale factor 
and $\epsilon = \epsilon_{\text{m},0} a^{-3}(t)$ is the energy density of 
dust, the energy density $\epsilon$ and pressure $p$ are never greater than
\[
\epsilon_{\text{max}} =  \frac{16m^{2}}{\hbar^{2}}, \qquad 
p_{\text{max}} = - \frac{16m^{2}}{\hbar^{2}},
\]
respectively. The energy density of the spin is negative but 
$\epsilon_{\text{eff}}$ is assumed to be positive. 
The ratio of the spin-induced term to the standard energy term is
\[
\frac{(8\pi G S)^{2}}{8 \pi G \rho} = 
\frac{8 \pi G \hbar^{2} \rho}{m^{2}}.
\]
With nucleons as dust particles, the above ratio is of the order 
$8 \times 10^{-55}\rho$ (where $\rho$ is expressed in the cgs units). 
Although the spin-spin contact interaction term appears to be negligibly
small even in what may be regarded now as superdense matter, it may play 
an essential role when the ratio approaches unity, i.e., in the earliest 
stages of evolution of the Universe. The spin-spin term produces something 
which may be called after Kopczynski the ``centrifugal force'' and which is 
able to prevent the occurrence of singularities in cosmology 
\cite{Kopczynski72,Kopczynski73}. 

Let us note now that effects of spin are dynamically equivalent to 
introducing into the model some additional non-interacting fluid for which 
the equation of state is
\begin{equation}
p_{\text{s}} = w_{\text{s}} \rho_{\text{s}}, 
\label{eq:4}
\end{equation}
where $w_{\text{s}} = 1$, $\rho_{\text{s}} = \rho_{\text{s},0} a^{-6}$, 
$\rho_{\text{s},0} = -(\hbar^{2}/16)n(0)$, just as for the Zeldovich 
stiff matter (or brane effects with dust on a brane with negative tension).

\section{The FRW model with spin on two-dimensional phase plane}

The dynamics of the Friedmann-Roberston-Walker (FRW) model with the 
Weyssenhoff fluid can be represented on the phase plane $(H,\epsilon)$, 
where $H = d(\ln a)/dt$ is Hubble's function, in the following way
\begin{subequations}
\label{eq:5}
\begin{align}
\frac{dH}{dt} &= - H^{2} - \frac{1}{6} (\epsilon_{\text{eff}} 
+ 3p_{\text{eff}}) + \frac{\Lambda}{3}
\label{eq:5a} \\
\frac{d\epsilon}{dt} &= -3H\epsilon
\label{eq:5b}
\end{align}
\end{subequations}
where $p_{\text{eff}}$ and $\epsilon_{\text{eff}}$ are parameterized by 
the energy density of dust (see formula~(\ref{eq:3})). Of course, 
system~(\ref{eq:5}) has the first integral
\begin{equation}
\epsilon_{\text{eff}} - 3H^{2} = \frac{3k}{a^{2}}- \Lambda,
\label{eq:6}
\end{equation}
where $k \in \{0, \pm 1\}$ is the curvature constant. Additionally, we have 
the conservation condition for non-interacting spin fluid with energy 
density $\epsilon$ 
\begin{equation}
\frac{d\epsilon_{\text{s}}}{dt} = -6H\epsilon_{\text{s}}.
\label{eq:7}
\end{equation}
The critical points of~(\ref{eq:5}) can be one of two admissible types: 
static if $H_{0} = 0$, $\epsilon_{0} \neq 0$, or 
non-static if $H_{0} \neq 0$ and $\epsilon_{0} = 0$. 
In the first case critical points lie on the intersection $H$-axis with the 
boundary condition $\epsilon_{\text{eff}} + 3p_{\text{eff}} = 0$ 
(the cosmological constant is formally included into $p_{\text{eff}}$ and 
$\rho_{\text{eff}}$ in the standard way $\rho_{\Lambda} = \Lambda$ and 
$p_{\Lambda} = -\Lambda$). The second type of critical points lie on the 
intersection of $\epsilon$-axis with the trajectory for flat model 
$\{\epsilon = 3H^{2}\}$. 

\begin{figure}
\includegraphics[angle=-90,width=0.75\textwidth]{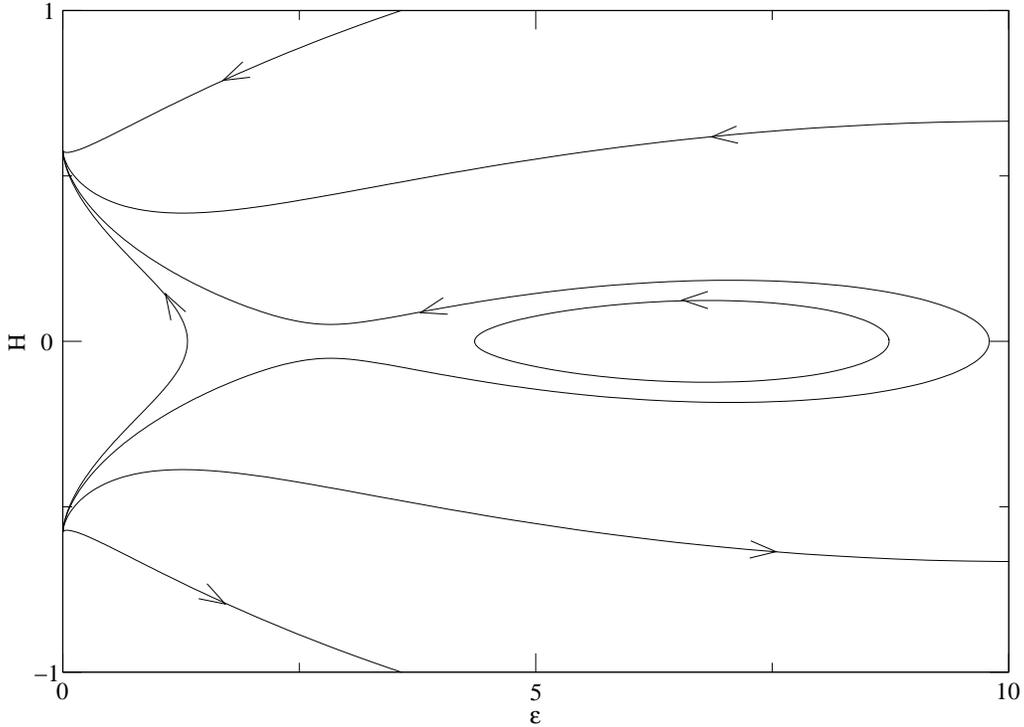}
\caption{The phase portrait of system~(\ref{eq:5}). There are the four 
critical points: stable node, unstable node, saddle point and center.
The presence of the center in the finite region indicates that the model 
is structurally unstable.}
\label{fig:1}
\end{figure}

The physically admissible domain for trajectories is
\[
\{ (H,\epsilon) \colon H \in \mathbb{R} \quad \text{and} \quad 
0 \leq \epsilon \leq \epsilon_{\text{max}}, \qquad
\epsilon_{\text{max}} \colon \epsilon_{\text{eff}}(\epsilon_{\text{max}}) 
= -\Lambda \}.
\]
The phase portrait of system~(\ref{eq:5}) with $\Lambda > 0$ is demonstrated 
on Fig.~\ref{fig:1}. 

\begin{figure}
\includegraphics[width=0.75\textwidth]{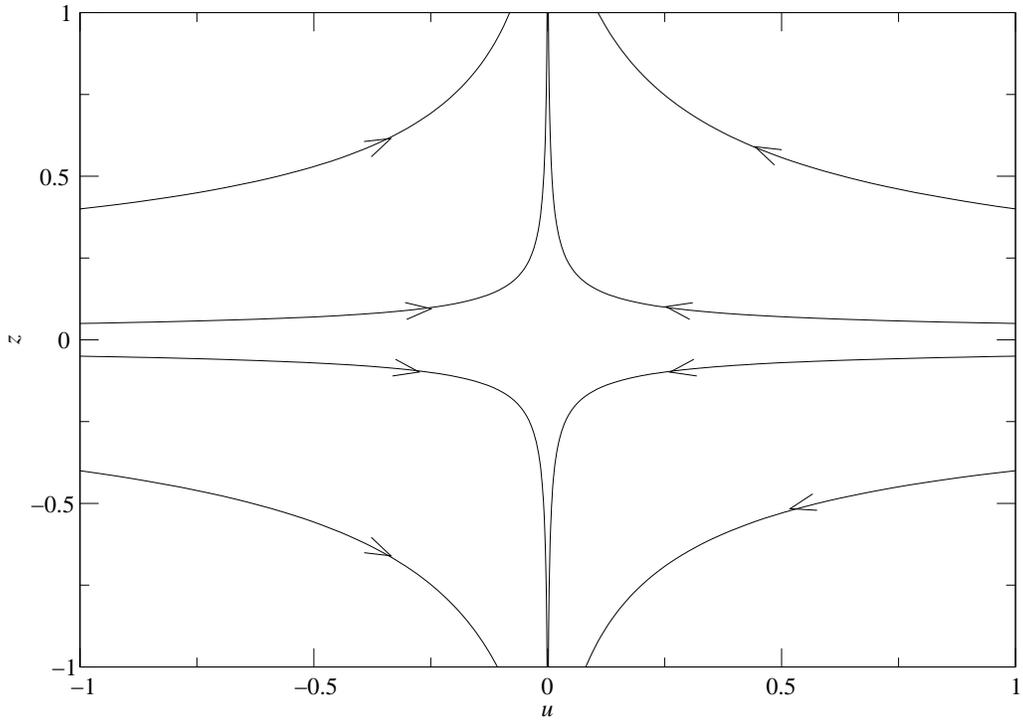}\\ \vspace{2cm}
\includegraphics[width=0.75\textwidth]{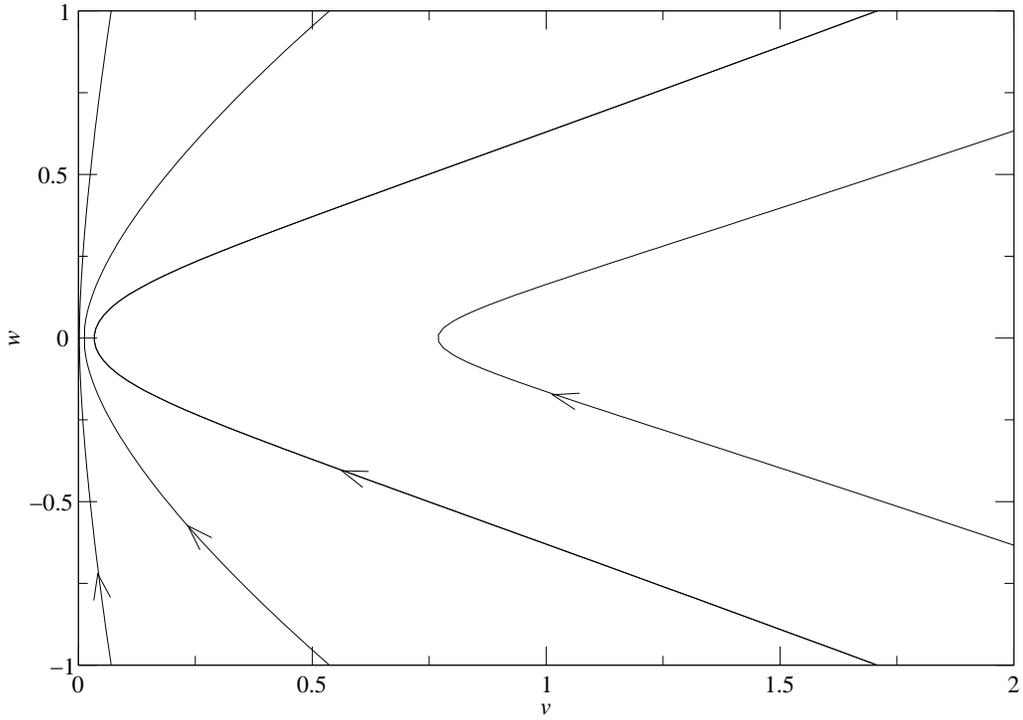}
\caption{The phase portraits of system~(\ref{eq:5}) in the projective 
coordinates $(z,u)$ and $(v,w)$. From the portrait in the former 
coordinates we can see that there is a saddle point in the infinity.}
\label{fig:2}
\end{figure}

The differences in the behavior of trajectories are manifest at high 
densities. Then the structure of dynamical behavior at infinity is modified. 
To illustrate the behavior of trajectories at infinity,
system~(\ref{eq:5}) is represented on Fig.~\ref{fig:2} in the projective 
coordinates 
\begin{subequations}
\label{eq:8}
\begin{align}
z &= \frac{1}{H}, \qquad u = \frac{\epsilon}{H}, & (z,u) &- \text{map} 
\label{eq:8a} \\
v &= \frac{1}{\epsilon}, \qquad w = \frac{H}{\epsilon}, & (v,w) &- \text{map}.
\end{align}
\end{subequations}
The two maps $(z,u)\colon z = 0, -\infty < u < \infty$ and $(v,w)\colon v = 0, 
-\infty < w < \infty$ cover the behavior of trajectories at the infinity 
circles $H = \infty$ and $\epsilon = \infty$.

\section{Dynamics of the model with spin fluid}

It is well known that the first integral of the FRW equation can be used 
to construct a Hamiltonian function. We take advantage of this feature in 
the considered model. The integration of equation (\ref{eq:5b}) gives 
$\rho = \rho_{0} a^{-3}$. Therefore the right-hand side of the Raychaudhuri 
equation (\ref{eq:5a}) can be expressed in terms of the scale factor $a(t)$ as
\begin{equation}
\label{eq:9}
\frac{d^{2} a}{dt^{2}}  = a \left[ - \frac{1}{6} \epsilon_{\text{m},0} 
a^{-3} + \frac{1}{6} \left( \frac{\hbar}{2} \right)^{2} a^{-6} 
+ \frac{\Lambda}{3} \right]. 
\end{equation}
Equation (\ref{eq:9}) can be rewritten in a form analogous to the Newton 
equation of motion in the one-dimensional configuration space 
$\{ a \colon a \in \mathbb{R}_{+} \}$
\begin{equation}
\label{eq:10}
\ddot{a} = - \frac{\partial V(a)}{\partial a}
\end{equation}
where the potential function
\begin{equation}
\label{eq:11}
V(a) = - \frac{1}{6} \epsilon_{\text{m},0} a^{-1} 
+ \frac{1}{24} \left( \frac{\hbar}{2} \right)^{2} a^{-4} 
- \frac{2}{3} \Lambda a^{2} + V_{0}
\end{equation}
and $V_{0} = \text{const}$. First integral (\ref{eq:6}) should correspond 
with the Hamiltonian integral of motion, hence
\[
V(a) + \frac{\dot{a}^{2}}{2} = V_{0} - \frac{k}{2}.
 \]
Now we construct the Hamiltonian function
\begin{equation}
\label{eq:12}
\mathcal{H} \equiv \frac{\dot{a}^{2}}{2} + V(a)
\end{equation}
and then the trajectories of the system lie on the energy level 
$\mathcal{H} \equiv E$, but if we choose $V_{0} = k/2$ then the physical 
trajectories lie on the zero-energy level $\mathcal{H} = E = 0$, which 
coincides with the form of first integral (\ref{eq:6}). Then finally 
we obtain
\begin{equation}
\label{eq:13}
V(a) = - \frac{1}{6} \epsilon_{\text{m},0} a^{-1} 
+ \frac{\hbar^{2}}{96} n(0) a^{-4} - \frac{1}{6} \Lambda a^2 + \frac{k}{2} 
\end{equation}
or in general
\begin{equation}
\label{eq:14}
V(a) = \frac{1}{6} \int^{a} (\epsilon_{\text{eff}} + 3p_{\text{eff}}) a da 
= - \frac{1}{6} \epsilon_{\text{eff}} a^{2} - \frac{\Lambda}{2} a 
+ \frac{k}{2}
\end{equation}
where
\[
\epsilon_{\text{eff}}(a) = \epsilon(a) + \epsilon_{\text{s},0} a^{-6}, \qquad 
\epsilon_{\text{s},0} = - \frac{1}{16} \hbar^{2} n(0),
\]
and $n(0)$ is an initial number of particles in the unit comoving volume. 

Finally we obtain dynamics reduced to the form of particle like problem in 
the one-dimensional potential
\begin{subequations}
\label{eq:15}
\begin{align}
\label{eq:15a}
\dot{x} &= y \\
\label{eq:15b} 
\dot{y} &= - \frac{\partial V}{\partial x}
\end{align}
\end{subequations}
with
\begin{equation}
\label{eq:15c}
V(x) = - \frac{1}{6}(\epsilon_{\text{m},0} x^{-1} 
+ \epsilon_{\text{s},0} x^{-4} - \epsilon_{\Lambda,0} x^{2} - 3k)
\end{equation}
where $x = a$, $y = \dot{a}$ and the above system has the first integral
\begin{equation}
\label{eq:16}
\frac{y^{2}}{2} + V(x) = 0. 
\end{equation}
In order, basic dynamical system~(\ref{eq:15}) is then rewritten as
\begin{subequations}
\label{eq:17}
\begin{align}
\label{eq:17a}
\frac{\dot{v}^{2}}{2} &= \frac{1}{2} \Omega_{k,0} + \frac{1}{2} 
\sum_{i} \Omega_{i,0} v^{-(1+3w_{i})} \\
\label{eq:17b}
\ddot{v} &= - \frac{1}{2} \sum_{i} \Omega_{i,0} (1+3w_i) v^{-(2+3w_i)}
\end{align}
\end{subequations}
where $v \equiv a/a_0$, $t \to T \equiv |H_0 |t$ is a new time variable 
denoted as dot in (\ref{eq:17}); $\Omega_{i} \in \{ \Omega_{\text{m},0},%
\Omega_{\text{s},0},\Omega_{\Lambda,0} \}$ are density parameters, 
$\Omega_{i} = \rho_{i} / \rho_{\text{cr}}$, 
$\Omega_{i,0} = \rho_{i,0} /3H_{0}^{2}$ 
and the subscript $0$ means that a quantity with this subscript is evaluated 
today (at time $t_{0}$); $\rho_{i} = \rho_{i,0} a^{-(1+w_{i})}$, where
$w_{\text{m}} = 0$ (dust), $w_{\text{s}} = 1$ (spin fluid), $w_{\Lambda} = -1$ 
(cosmological constant), and $p_{i} = w_{i} \rho_{i}$.

The representation of dynamics as a one-dimensional Hamiltonian flow allows 
to make the classification of possible evolution paths in the configuration 
space which is complementary to phase diagrams. It also makes simpler to 
discuss the physical content of the model. Finally, the construction of 
the Hamiltonian allows to study quantum cosmology models with spin fluid 
in full analogy to what is usually done in general relativity.

From equation~(\ref{eq:14}) we can observe that trajectories are integrable 
in quadratures. Namely, from the Hamiltonian constraint $\mathcal{H} = E = 0$ 
we obtain the integral
\begin{equation}
\label{eq:18}
t-t_{0} = \int_{a_{0}}^{a} \frac{da}{\sqrt{-2V(a)}}.
\end{equation}
For some specific forms of potential function~(\ref{eq:15c}) we can obtain 
exact solutions. 

It is possible to make the classification of qualitative evolution paths 
by analyzing the characteristic curve which represents the boundary equation 
of the domain admissible for motion. For this purpose we consider the 
equation of zero velocity, $\dot{a} = 0$ which constitutes the boundary 
$\mathcal{M} = \{a\colon V(a) = 0\}$.

\begin{figure}
\includegraphics[width=0.75\textwidth]{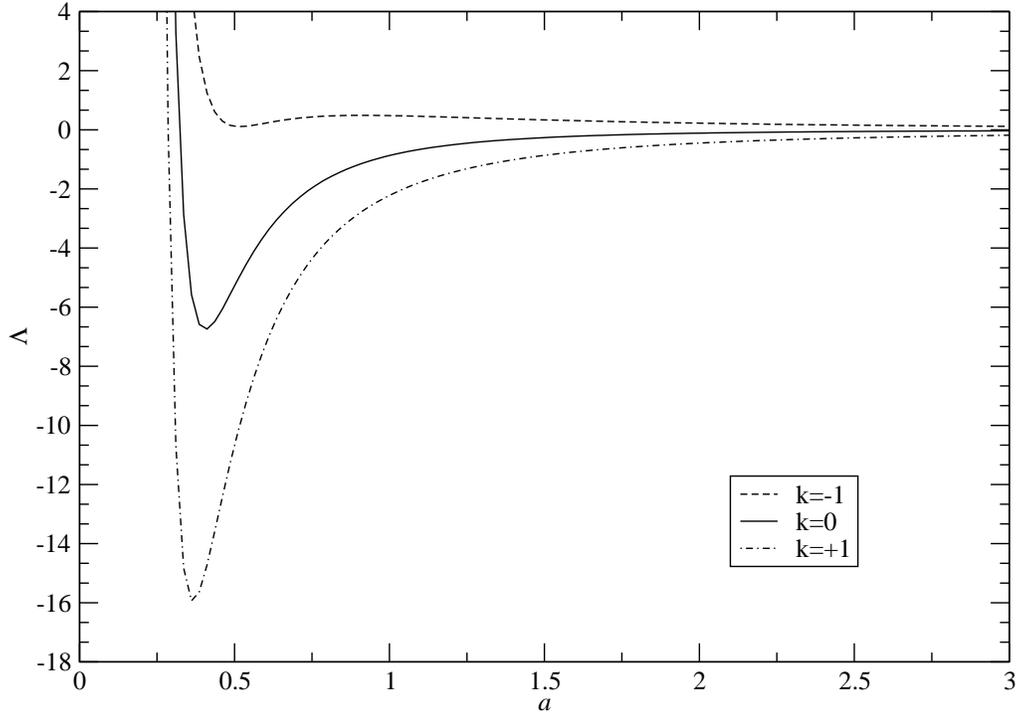}
\caption{The dependence of $\Lambda$ on $a$.}
\label{fig:3}
\end{figure}

From equation~(\ref{eq:15c}) the cosmological constant can be expressed as 
a function of $x$ as follows
\begin{equation}
\label{eq:19}
\Lambda(x) = x^{-2} (\epsilon_{\text{m},0} x^{-1} 
- \epsilon_{\text{s},0} x^{-4} - 3k).
\end{equation}
The plot of $\Lambda(x)$ for different $k$ is shown in Fig.~\ref{fig:3}. 
The domain under the curve $\Lambda(x)$ is non-physical, and we 
consider the evolution path as a level of $\Lambda = \text{const}$ 
and we classify all evolution models with respect to their quantitative 
properties of dynamics. For negative $\Lambda$ there are oscillating 
solutions without a singularity.

The next advantage of representing dynamics in terms of Hamiltonian is 
possibility to discuss how trajectories along which the acceleration 
condition $\ddot{a} = -dV/da > 0$ is satisfied are distributed in the phase 
space. One can easily observe this phenomenon from the geometry of the 
potential function. In the phase plane the area of acceleration is determined 
by $\dot{y} > 0$ or by the condition that $V(x)$ is a decreasing function of 
its parameter
\begin{equation}
\label{eq:20}
- \frac{1}{2} \Omega_{\text{m},0} v^{-2} 
- 2 \Omega_{\text{s},0} v^{-5} + \Omega_{\Lambda,0} v > 0
\end{equation}
where
\[
v = \frac{a}{a_{0}}, \qquad \Omega_{\text{s},0} \le 0, \qquad 
\Omega_{k,0} = 0.
\]

Independent observations of supernovae type Ia made by the Supernovae 
Cosmology Project and High-z Survey Team indicate that our Universe is 
presently accelerating. There is a fundamental problem in explaining this 
acceleration. If we introduce the cosmological constant and assume 
$\Omega_{k,0} = 0$ (the universe is flat) then the best fit model is 
$\Omega_{\Lambda,0} = 0.72$, $\Omega_{\text{m},0} = 0.28$. 

The effects of spin fluid cannot dominate the matter contributions during 
the whole evolution of the universe. But we argue that the effects of spin 
degree of freedom are important in early universe. In any case they should 
be smaller than or comparable with the matter contribution because 
\[
\frac{H^2}{H_{0}^{2}} = \Omega_{\mathrm{m,0}} x^{-3(1+\gamma} 
+ \Omega_{\mathrm{s},0}x^{-6} \ge 0.
\]

The formalism presented gives us a natural base to discuss this problem 
for the FRW model with the Weyssenhoff fluid. It is convenient to introduce 
a new variable $z = v^{-3}$ , then inequality~(\ref{eq:20}) reduces to the 
quadratic inequality
\[
- \frac{1}{2} \Omega_{\text{m},0} z + 2|\Omega_{\text{s},0}|z^{2} 
+ \Omega_{\Lambda,0} > 0.
\]
Therefore the Universe is presently accelerating provided that
\[
- \frac{1}{2} \Omega_{\text{m},0} + 2|\Omega_{\text{s},0}| 
+ \Omega_{\Lambda,0} > 0.
\]
For instance, for $\Omega_{\text{m},0} = 0.3$, $\Omega_{\Lambda,0} = 0$, 
$|\Omega_{\text{s},0}| > \frac{1}{4}\Omega_{\text{m},0}$ is required. 
Therefore, present experimental estimates based on baryons in clusters 
on one hand giving $\Omega_{\text{m},0} \sim 0.3$ and the location of the 
first peak in the cosmological microwave background detected by Boomerang 
and Maxima suggesting an early, filled by spin matter, flat universe 
on the other hand, imply that our Universe without a cosmological term would 
be presently accelerating only if
\[
|\Omega_{\text{s},0}| \gtrsim \frac{1}{3}.
\]
The required value of $|\Omega_{\text{s},0}|$ seems to be unrealistic 
(see the next section) and therefore the cosmological constant term is still 
needed to explain the present acceleration of the Universe. As we will 
see in the subsequent analysis the effect of spin fluid is negligible in 
the present epoch. Therefore the spin is an additional factor that can 
influence the dynamics of the early universe together with for example the 
cosmological constant.

\section{Redshift-magnitude relation for the model with spin fluid}

Cosmic distance measures like the luminosity distance depend sensitively 
on the spatial geometry (curvature) and the dynamics. Therefore, the 
luminosity depends on the present density parameters of different components 
and their form of the equation of state. For this reason the redshift-magnitude 
relation for distant galaxies is proposed as a potential test for the FRW 
model with spin fluid.

Let us consider an observer located at $r = 0$ at the moment $t = t_{0}$ 
which receives a light ray emitted at $t = t_{1}$ from the source of the 
absolute luminosity $L$ located at the radial distance $r_{1}$. The redshift 
$z$ of the source is related to the scale factor at the two moments of 
evolution by $1 + z = a(t_{0})/a(t_{1}) = a_{0}/a$. If the apparent 
luminosity of the source as measured by the observer is $l$ then the 
luminosity distance $d_{L}$ of the source is defined by the relation
\[
l = \frac{L}{4 \pi d_{L}^{2}}
\]
where $d_{L} = (1 + z)a_{0} r_{1}$.

The important test to verify whether the spin fluid FRW model may represent 
dark energy is the comparison with the supernova type Ia data. In order to 
do so, we evaluate the luminosity distance in the flat FRW model with spin 
fluid. In such a model, the luminosity distance reads
\[
d_{L}(z) = \frac{1 + z}{H_0} \int_{0}^{z} 
\frac{dz'}{\sqrt{\Omega_{\text{m},0}(1+z')^{3} 
+ \Omega_{\text{s},0}(1+z')^{6} + \Omega_{\Lambda,0}}}.
\]
From this expression the following relation between the apparent magnitude 
$m$ and absolute magnitude $M$ is obtained
\[
m - \mathcal{M} = 5\log\left[ (1 + z) \int_{0}^{z} 
\frac{dz'}{\sqrt{\Omega_{\text{m},0}(1+z')^{3} 
+ \Omega_{\text{s},0}(1+z')^{6} + \Omega_{\Lambda,0}}}\right]
\]
where $\mathcal{M} = M - 5\log H_{0} + 25$.

In order to compare with the supernova data, we compute the distance 
modulus
\[
\mu_{0} = 5\log(d_{L}) + 25
\]
where $d_{L}$ is in Mps. The goodness of fit is characterized by 
the parameter
\begin{equation}
\chi^{2}=\sum_{i} \frac{|\mu_{0,i}^{0}-\mu_{0,i}^{t}|}{\sigma_{\mu 0,i}^{2} 
+ \sigma_{\mu z,i}^{2}}.
\label{eq:21}
\end{equation}
In expression (\ref{eq:21}), $\mu_{0,i}^{0}$ is the measured value, 
$\mu_{0,i}^{t}$ is the value calculated in the model described above, 
$\sigma_{\mu 0,i}^{2}$ is the measurement error, $\sigma_{\mu z,i}^{2}$ 
is the dispersion in the distance modulus due to peculiar velocities of 
galaxies.

\section{The model with spin fluid tested by supernovae}

We test the model with spin fluid using sample A of the Perlmutter SN Ia data. 
In order to avoid any possible selection effects we work with this full 
sample of 60 supernovae. In the statistical analysis we use the maximum 
likelihood method with the marginalization procedure \cite{Riess98}.

First we estimated $\mathcal{M}$ from the full sample of 60 supernovae. For 
the flat model we obtained $\mathcal{M} = -3.39$.

\begin{figure}
\includegraphics[width=0.75\textwidth]{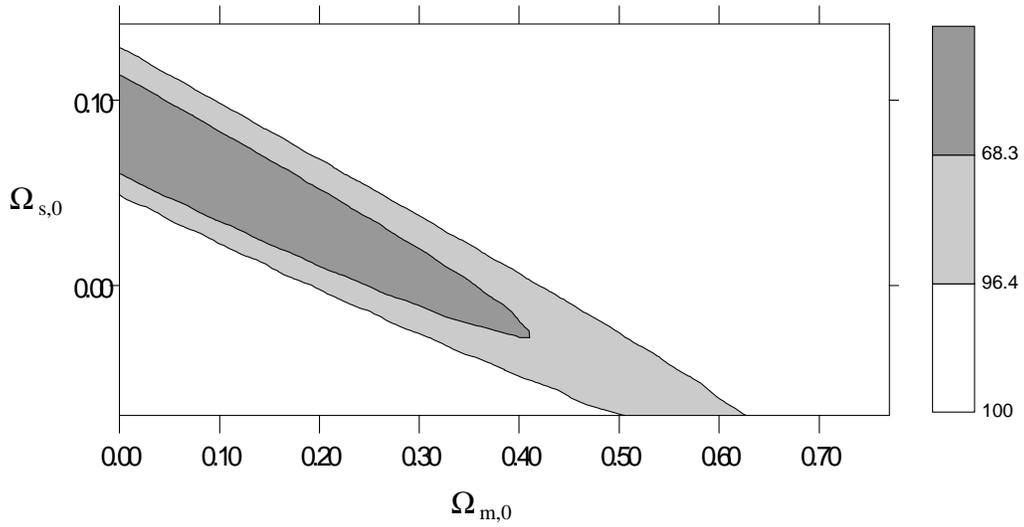}
\caption{Confidence levels on the plane 
$(\Omega_{\text{m},0} , \Omega_{\text{s},0})$ minimized over $\mathcal{M}$ 
for the flat model, and with 
$\Omega_{\Lambda,0}=1-\Omega_{\text{m},0}-\Omega_{k,0}-\Omega_{\text{s},0}$.
The figure shows the ellipses of the preferred value of 
$\Omega_{\text{m},0}$ and $\Omega_{\Lambda,0}$. The results prefer the 
positive value of $\Omega_{\text{s},0}$, while the negative values are 
allowed (i.e., spin fluid can exist).}
\label{fig:4}
\end{figure}

The result of the statistical analysis is presented in the figures. 
Figure~\ref{fig:4} illustrates the confidence level as a function of 
$\Omega_{\text{m},0}$, $\Omega_{\text{s},0}$ for the flat model 
($\Omega_{k,0}=0$) minimized over $\mathcal{M}$ with 
$\Omega_{\Lambda,0}=1-\Omega_{\text{m},0}-\Omega_{k,0}-\Omega_{\text{s},0}$. 

\begin{figure}
\includegraphics[width=0.75\textwidth]{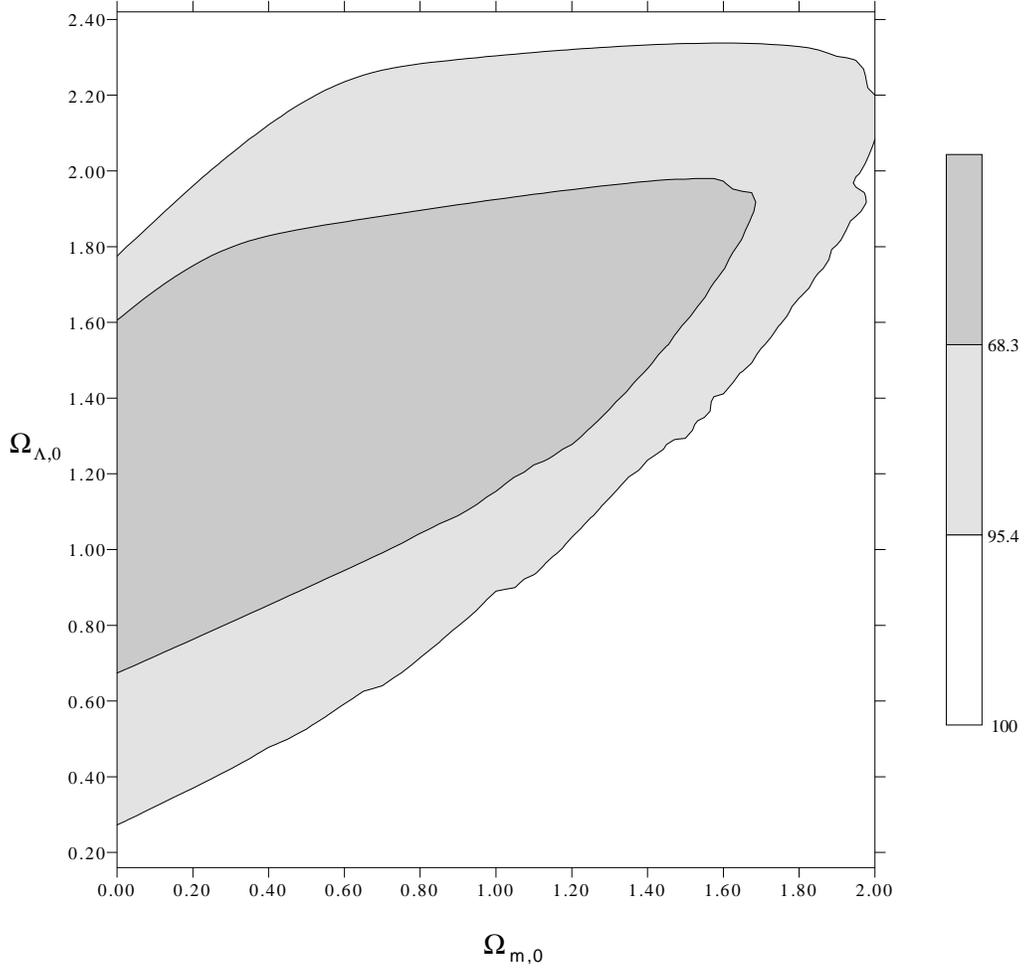}
\caption{Confidence levels on the plane 
$(\Omega_{\text{m},0} , \Omega_{\Lambda,0})$ minimized over $\mathcal{M}$, 
and with 
$\Omega_{\Lambda,0}=1-\Omega_{\text{m},0}-\Omega_{k,0}-\Omega_{\text{s},0}$.
The figure shows the ellipses of the preferred values of 
$\Omega_{\text{m},0}$ and $\Omega_{\Lambda,0}$.}
\label{fig:5}
\end{figure}

In the case of $\Omega_{k,0} \ne 0$, the preferred values of the pairs 
$(\Omega_{\text{m},0} ,\Omega_{\Lambda,0})$ are shown in a standard way 
after minimizing over $\mathcal{M}$ on Fig.~\ref{fig:5}. One can see that 
the non-zero cosmological constant is still required. 

\begin{figure}
\includegraphics[width=0.75\textwidth]{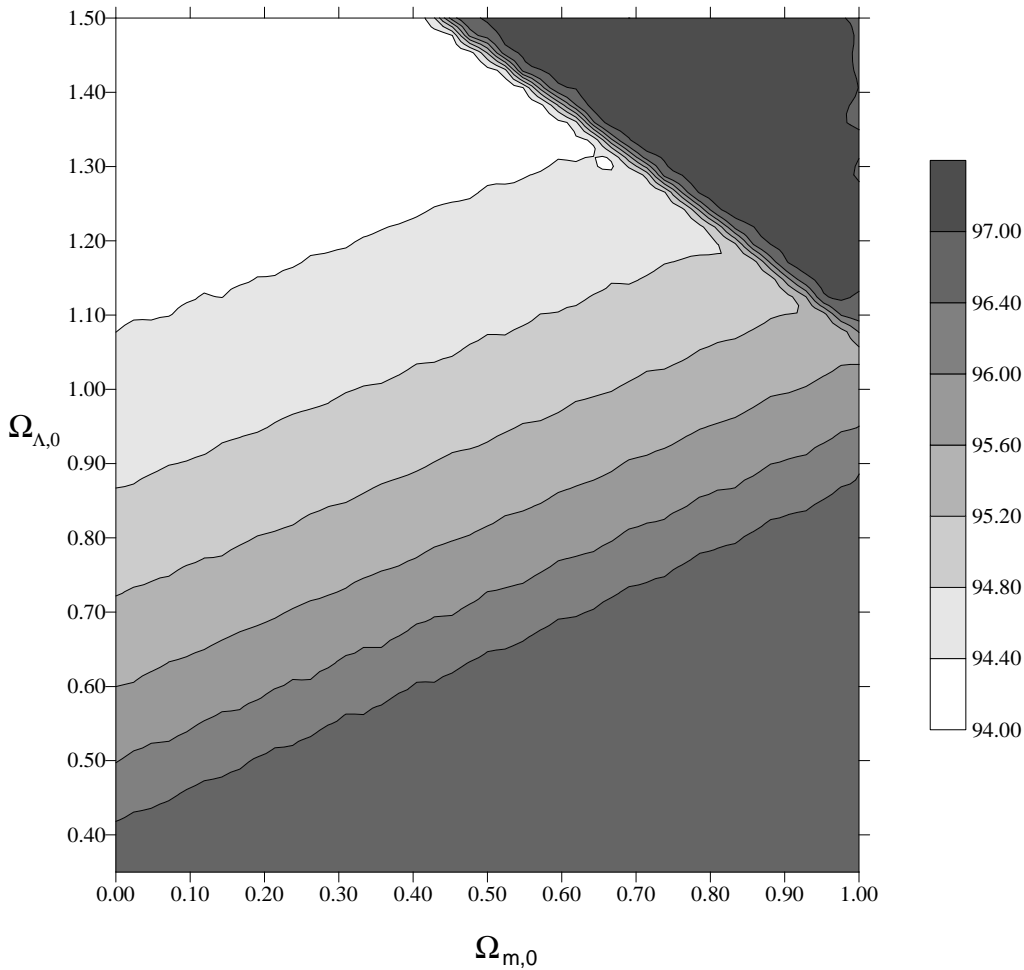}
\caption{Levels of constant 
$\chi^{2}$ on the plane $(\Omega_{\text{m},0},\Omega_{\Lambda,0})$ 
marginalized over $\mathcal{M}$, and with
$\Omega_{\Lambda,0}=1-\Omega_{\text{m},0}-\Omega_{k,0}-\Omega_{\text{s},0}$
The figure shows the preferred value of 
$\Omega_{\text{m},0}$ and $\Omega_{\Lambda,0}$.}
\label{fig:6}
\end{figure}

Applying the marginalization procedure over $\Omega_{k,0} = (-1, 1)$, 
$\Omega_{\text{s},0} < 0$, $\mathcal{M} \in (-4 , -3)$ we find the lowest 
value of $\chi^2$ for each pair of values 
$(\Omega_{\text{m},0},\Omega_{\Lambda,0})$ as shown on Fig.~\ref{fig:6}.
This figure shows the favored region of values of 
$(\Omega_{\text{m},0},\Omega_{\Lambda,0})$ (best fit).

\begin{figure}
\includegraphics[width=0.75\textwidth]{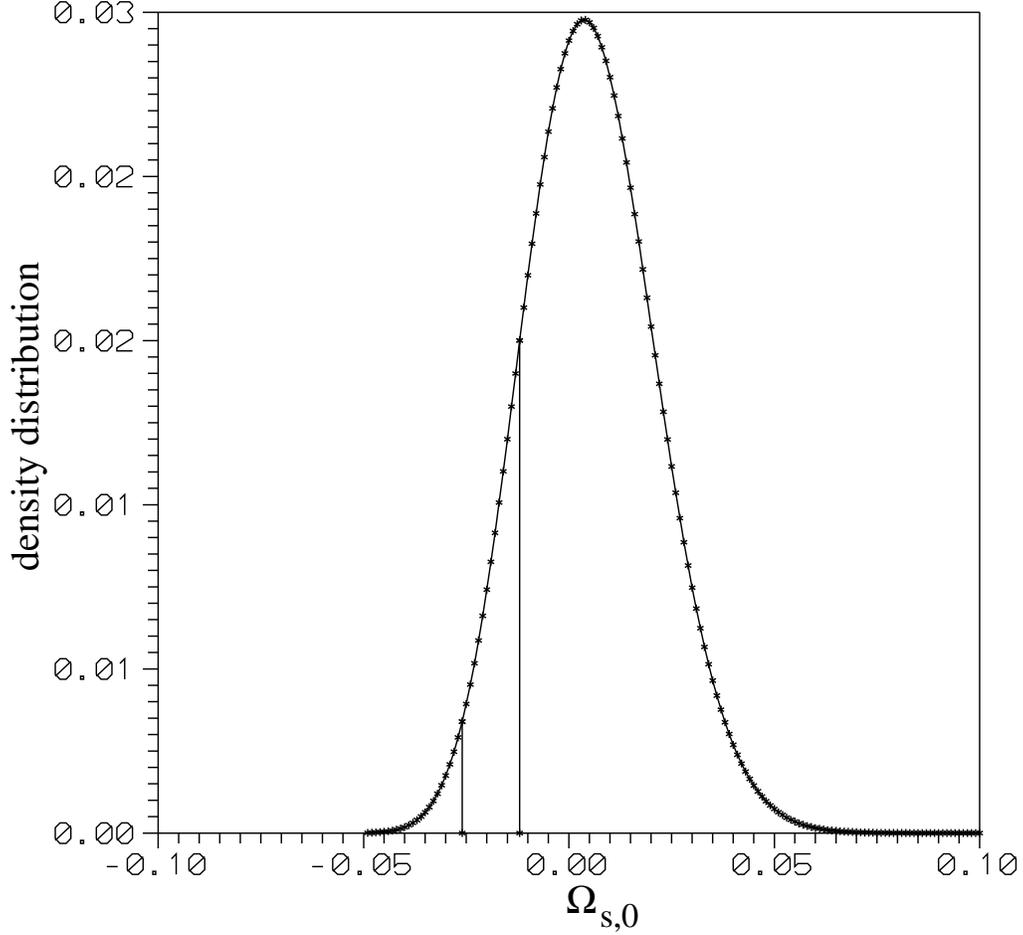}
\caption{The density distribution for 
$\Omega_{\text{s},0}$ in the model with spin effects 
($\Omega_{\text{s},0} < 0$) and brane effects 
($\Omega_{\text{s},0} > 0$). We obtain that $\Omega_{\text{s},0} > -0.012$ 
at the confidence level $1\sigma$ and $\Omega_{\text{s},0} > -0.026$ on 
the confidence level $2\sigma$. For simplicity, we only mark limits for the 
spin side.}
\label{fig:7}
\end{figure}

Figure~\ref{fig:7} shows the density distribution for $\Omega_{\text{s},0}$ 
in the flat model with $\Omega_{\text{m},0} = 0.3$. This distribution is 
obtained from the marginalization over $\mathcal{M}$. The positive values 
of $\Omega_{\text{s},0}$ can be interpreted as the brane effects with dust 
on the brane \cite{Szydlowski02}. One can see that 
$\Omega_{\text{s},0} > -0.012$ at the confidence level $1\sigma$ and 
$\Omega_{\text{s},0} > -0.026$ at the confidence level $2\sigma$. The limit 
value of $\Omega_{\text{s},0} = -0.012$ is used in our further analysis.

\begin{figure}
\includegraphics[width=0.75\textwidth]{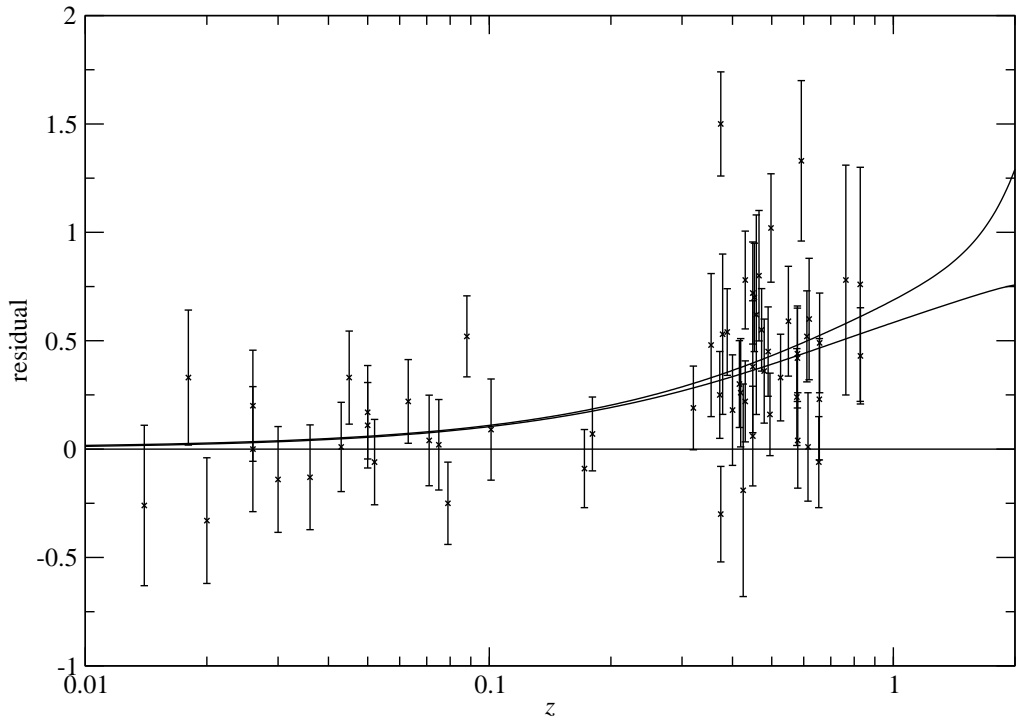}
\caption{For 
$\Omega_{\text{m},0}=0.3$, residuals between the 
Einstein-de Sitter model and three cases: 
the Einstein-de Sitter itself (zero line), the $\Lambda$CDM flat model 
(middle curve), and the flat model with spin matter for
$\Omega_{\text{s},0} = -0.012$ (highest curve).}
\label{fig:8}
\end{figure}

In Fig.~\ref{fig:8} we present the plot of residuals of redshift-magnitude 
relationship for the supernovae data. With the increasing impact of spin 
(lower $\Omega_{\text{s},0}$) the high-redshift supernovae should be fainter 
than the expected by the $\Lambda$CDM model. For $\Omega_{\text{s},0} = -0.012$ 
the difference between the spin model and the $\Lambda$CDM model should be 
detectable for $z > 1$. 

\begin{figure}
\includegraphics[width=0.75\textwidth]{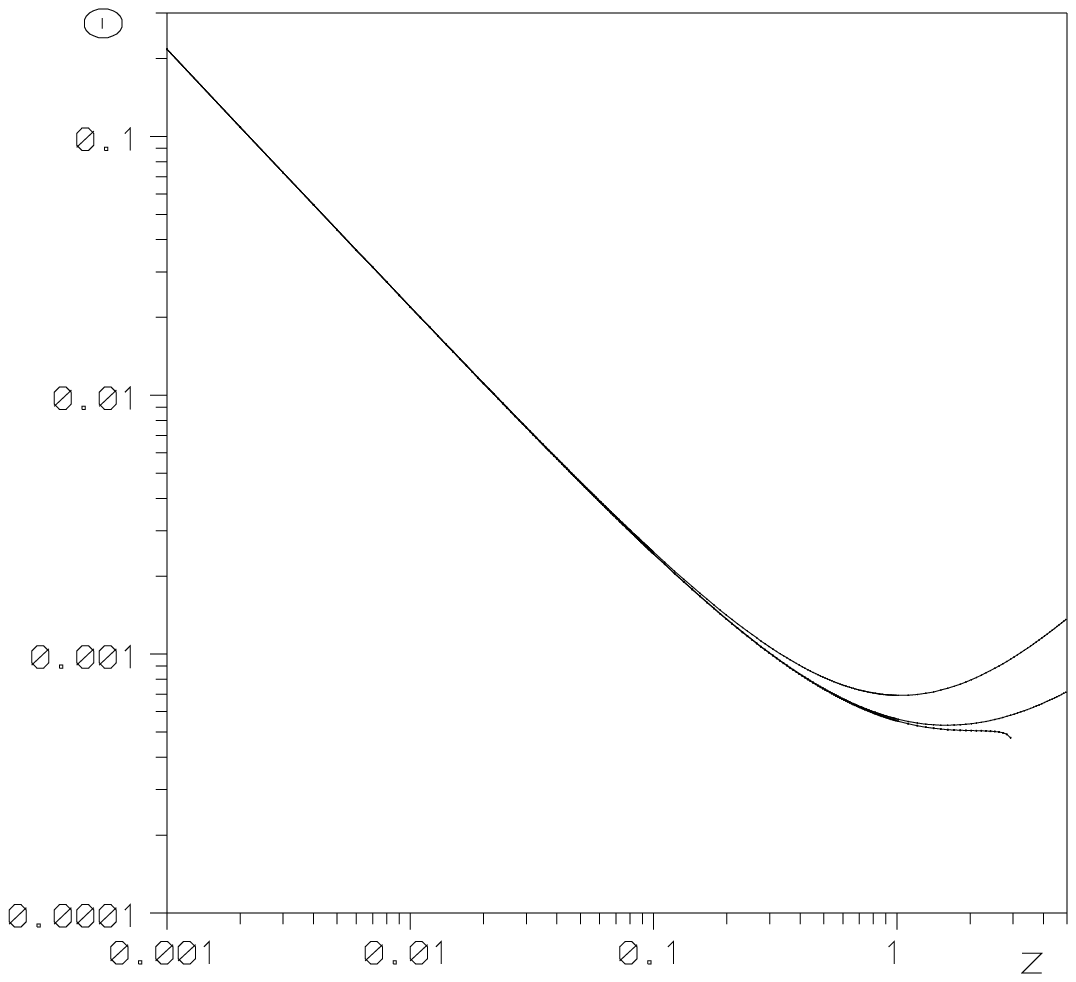}
\caption{The angular diameter $\theta$ for the flat model with spin for 
$\Omega_{\text{m},0} = 0.3$ and $\Omega_{\text{s},0} = 0.1,0,-0.005$ 
(top, middle, bottom, respectively). The minima for these cases are $1.04$, 
$1.605$, no minimum, respectively. The spin causes the minimum to move 
right (toward higher $z$) and the minimum value of $\theta$ decreases.}
\label{fig:9}
\end{figure}

The angular diameter of a galaxy is defined as
\[
\theta = \frac{d(z + 1)^{2}}{d_{L}}
\]
where $d$ is the linear size of a galaxy. In the standard cosmology the flat 
dust-filled universe $\theta$ has the minimum value $z_{\text{min}} = 5/4$. 
From Fig.~\ref{fig:9} we can see that, if the spin matter is present, 
its influence on the predicted $\theta$ is weak. Theoretically it is possible 
to test values of $\Omega_{\text{s},0}$ from the angular diameter minimum 
value test. But because of evolutionary effects and observational difficulties 
the predicted differences are too small to be detected.

\begin{figure}
\includegraphics[width=0.75\textwidth]{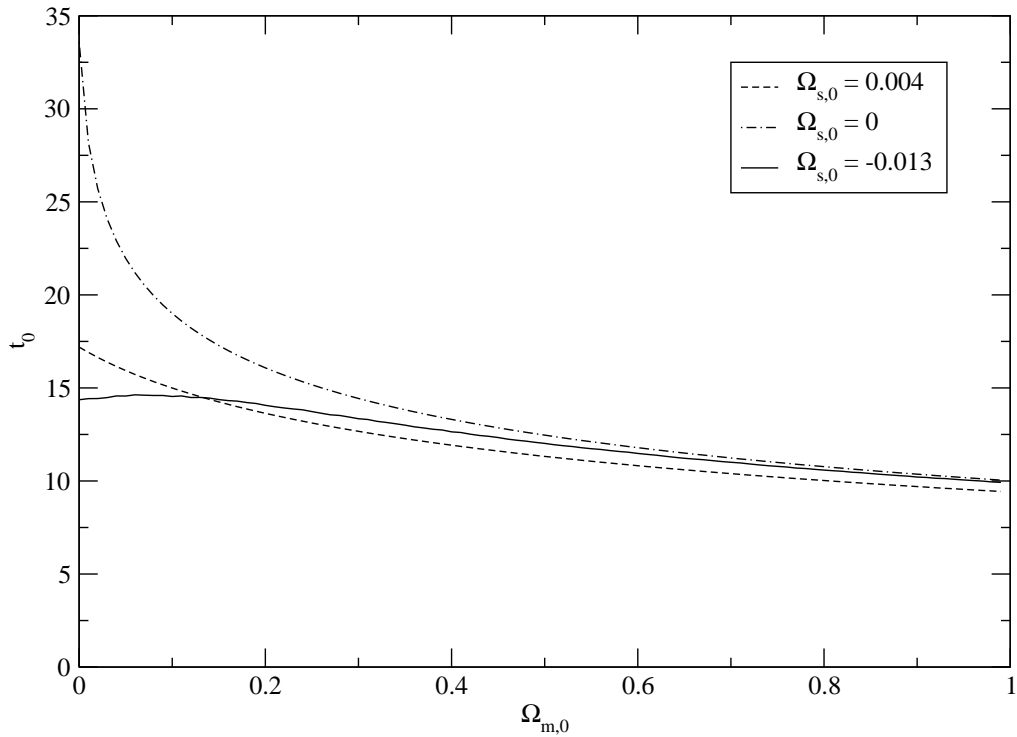}
\caption{The dependence of the age of the universe 
$t_{0}$ (in units of $10^{9}$~yr) on the parameter 
$\Omega_{\text{m},0}$ for the flat model with spin fluid.}
\label{fig:10}
\end{figure}

Now let us briefly discuss the effect of spin fluid on the age of the Universe 
which is given by
\[
t_{0} = \frac{1}{3 H_{0} \sqrt{\Omega_{\Lambda,0}}} 
\ln \frac{2\Omega_{\text{m},0} + 2\Omega_{\Lambda,0} 
+ \sqrt{\Omega_{\Lambda,0}}}{\sqrt{\Omega_{\text{m},0}^{2} 
- 4\Omega_{\text{s},0}\Omega_{\Lambda,0}}}.
\]
In Fig.~\ref{fig:10} we plot the age of the Universe in Gyr for the 
flat model for different $\Omega_{\text{s},0}$. 
Taking $\Omega_{\text{m},0}=0.3$ we obtain for $\Omega_{\text{s},0} = -0.013$: 
$t_{0} = 13.347$, and for $\Omega_{\text{s},0} = 0$: $t_{0} = 14.436$.
We can see that the spin fluid lowers significantly the age of the Universe. 
The cosmological constant is still needed to explain the problem of the age 
of the Universe.

\section{CMB peaks in the model with spin fluid}

The cosmic microwave background (CMB) peaks arise from acoustic oscillations 
of the primeval plasma. Physically these oscillations represent hot and cold 
spots. Thus, a wave that has a density maximum at the time of last 
scattering corresponds to a peak in the power spectrum. In the Legendre 
multipole space this corresponds to the angle subtended by the sound horizon 
at the last scattering. Higher harmonics of the principal oscillations which 
have oscillated more than once correspond to secondary peaks.

For our end it is very important that the locations of these peaks are very 
sensitive to the variations in the model parameters. Therefore, it can 
serve as a sensitive probe to constrain the cosmological parameters and 
discriminate among various models.

The locations of the peaks are set by the acoustic scale $l_{A}$ which can 
be defined as the angle $\theta_{A}$ subtended by the sound horizon at the 
last scattering surface. The acoustic scale $l_{A} = \pi/\theta_{A}$ in 
the flat model is given by
\[
l_{A} = \pi \frac{\int_{0}^{z_{\text{dec}}} \frac{dz'}{H(z')}}
{\int_{z_{\text{dec}}}^{\infty} c_{s} \frac{dz'}{H(z')}}
\]
where
\[
H(z) = H_{0} \sqrt{\Omega_{\text{r},0} (1 + z)^{4} 
+ \Omega_{\text{s},0} (1 + z)^{6} + \Omega_{\text{m},0} (1 + z)^{3} 
+ \Omega_{\Lambda,0}}
\]
and $c_{s}$ is the speed of sound in the plasma (we assume additionally the 
presence of radiation in the model). The sound velocity can be calculated 
from the formula
\[
c_{s}^{2} \equiv \frac{d p_{\text{eff}}}{d \rho_{\text{eff}}} 
= \frac{6\Omega_{\text{s},0}(1+z)^{3} + \frac{4}{3} \Omega_{\gamma,0}(1+z)}
{3 \Omega_{\text{m},0} + 4 \Omega_{\gamma,0}(1+z) 
+ 6 \Omega_{\text{s},0}(1+z)^{3}}.
\]
In the model of primeval plasma, there is a simple relation
\[
l_{m} \approx l_{A}(m - \phi_{m})
\]
between the location of $m$-th peak and the acoustic scale 
\cite{Doran01,Hu01}. The prior assumptions in our calculations are 
as follows $\Omega_{\text{r},0} = 9.89 \times 10^{-5}$, 
$\Omega_{\text{b},0} = 0.05$, and the spectral index for initial 
density perturbations $n = 1$, and $h = 0.65$.

The phase shift is caused by the prerecombination physics (plasma driving 
effect) and, hence, has no significant contribution from the term containing 
spin in that epoch. Because of above assumptions the phase shift $\phi_{m}$ 
can be taken from standard cosmology \cite{Hu01} 
\[
\phi_{m} \approx 0.267 \left[ \frac{r(z_{\text{dec}})}{0.3} \right]^{0.1}
\]
where $\Omega_{\text{b},0} h^{2} = 0.02$, $r(z_{\text{dec}}) \equiv 
\rho_{\text{r}}(z_{\text{dec}})/\rho_{\text{m}}(z_{\text{dec}}) 
= \Omega_{\text{r},0}(1 + z_{\text{dec}})/\Omega_{\text{m},0}$, 
$\Omega_{\text{r},0} = \Omega_{\gamma,0} + \Omega_{\nu,0}$, 
$\Omega_{\gamma,0} = 2.48h^{-2} \times 10^{-5}$, 
$\Omega_{\nu,0} = 1.7h^{-2} \times 10^{-5}$, $r(z_{\text{dec}})$ 
is the ratio of radiation to matter densities at the surface of last 
scattering. 

The influence of spin on the location of the peaks is to shift them towards 
higher values of $l$. For example, for $\Omega_{\text{m},0} = 0.3$, 
$\Omega_{\text{b},0} = 0.05$, $h = 0.65$, the different choices of 
$\Omega_{\text{s},0}$ yield the following
\begin{align*}
\Omega_{\text{s},0} &= 10^{-10} \colon & l_{\text{peak},1} &= 186, 
& l_{\text{peak},2} = 441 \\
\Omega_{\text{s},0} &= 1.4 \times 10^{-10} \colon & l_{\text{peak},1} &= 223, 
& l_{\text{peak},2} = 530
\end{align*}
On the other hand from the Boomerang observations \cite{deBernardis02} 
we obtain $l_{\text{peak},1} = 200-223$, $l_{\text{peak},2} = 509-561$. 
Therefore, uncertainties in values $l_{\text{peak}}$ can be used in 
constraining cosmology with spin fluid, namely
\[
10^{-10} \leq |\Omega_{\text{s},0}| \leq 1.4 \times 10^{-10}
\]
from the location of the first peak.

\section{BBN in the model with spin fluid}

We consider the non-relativistic matter with density $\rho$. In this 
case the spin effects scale like $-(1+z)^6$. It is clear that such a 
term can lead to accelerated expansion and the detailed analysis of 
supernovae Ia data requires it to be very subdominant today. 

However, going back in time to the big bang nucleosynthesis (BBN) period, 
the term $\Omega_{\text{s},0}= -0.012$ would be dominant at redshift 
$z \gtrsim 2$. In such a model, radiation domination would never occur and 
all the BBN predictions would be lost. In practice, the preferred value 
obtained from SN Ia data gives the term $\rho_{\text{s}}^{2}$ which is 
far too large to be compatible with the BBN which is very well tested area 
of cosmology and does not allow for significant deviation from the 
standard expansion law, apart from very early times before the onset 
of the BBN. The consistency with the BBN seems to be crucial issue 
in spin fluid cosmology \cite{Arkani-Hamed99,Binetruy00}.
For this reason, not to suffer from the contradiction with the BBN, 
the contribution of spin fluid $\Omega_{\text{s},0}$ cannot 
dominate over the standard radiation term before the onset of BBN, 
i.e., for $z \cong 10^8$
\[
- \Omega_{\text{s},0}(1+z)^6 < \Omega_{\text{r},0}(1+z)^4 
\quad \text{and} \quad - \Omega_{\text{s},0}<10^{-20}.
\]

Therefore, the term $\Omega_{\text{s},0}(1+z)^6$, describing the spin effects, 
is constrained by the BBN because it requires the change of expansion rate 
due to this term to be sufficiently small, so that an acceptable helium-4 
abundance is produced.

\section{Conclusions}

We discussed dynamics of the FRW model with spin fluid called the 
Weyssenhoff fluid. We showed that the effect of spin fluid is equivalent 
to fictitious fluid with equation like the Zeldovich stiff matter and 
negative energy density.

The dynamics of these models was analyzed on the two-dimensional phase 
plane and the Hamiltonian formalism was adopted to analyze all evolutional 
paths in the configuration space.

The dynamics determines the luminosity distance relationship which is used 
to test the models using the supernovae type Ia observations. It is shown 
that the presence of spin fluid matter has no influence on the present 
acceleration rate of the universe, and it is not an alternative to the 
cosmological constant description of dark energy.

We showed that the difference between the $\Lambda$CDM model and the model 
with spin fluid would be detectable if the content of spin fluid matter was 
sufficiently large, e.g., $\Omega_{\text{s},0} = -0.012$. This is possible 
because supernovae with redshift $z > 1$ should be fainter in the model 
with spin fluid than in the $\Lambda$CDM model.

The detailed statistical analysis of supernovae data of sample A (analysis 
of confidence levels, levels of constant $\chi^{2}$ and density distribution 
of probability) was performed. From this analysis we obtained the limit of 
the density parameter for spin fluid $\Omega_{\text{s},0}$. At the confidence
level of $1\sigma$ it is equal to $-0.012$.

We also applied the test of the minimum of the angular size of galaxies and 
showed that it is sensitive to the amount of spin fluid matter but difficult 
to detect. We found that the the presence of spin fluid matter lowers the 
age of the Universe.

Moreover we demonstrated that the stronger limit can be obtained from 
the CMB peak locations using the Boomerang data. From the uncertainties of 
the location of first peak we obtained a small interval for the values of spin 
fluid parameter $-1.4 \cdot 10^{-10} \leq \Omega_{\text{s},0} \leq -10^{-10}$.
 
We find the formal analogy between the model with spin fluid and the 
brane model with dust on a brane. In both cases the dynamics equations are 
formally equivalent. In the early Universe these two kinds of matter scaled 
according the same law. For this reason we also considered the positive 
values of the spin fluid parameter $\Omega_{\text{s},0}$. From the statistical 
analysis we observed that a positive value of spin fluid parameter (brane) 
is most probable, nevertheless a negative value of this parameter is 
statistically allowed.

In the near future new high-redshift SN Ia observations will bring on better 
data. We expect that they will allow us to obtain the estimation of 
$\Omega_{\text{s},0}$ with a significantly lower error and to restrict the 
limit for spin fluid. In such a case the estimation of $\Omega_{\text{s},0}$ 
using SN Ia data will be worth reconsideration because this method does not 
depend on model assumptions, while the BBN limit on $\Omega_{\text{s},0}$ is 
strongly model dependent.

\acknowledgments{
The paper was supported by KBN grant No. 1 P03D 003 26. 
We thank dr W. God{\l}owski and W. Czaja for comments.}

\end{document}